\documentclass[12pt]{revtex4-2}
\usepackage{rotating}
\usepackage{makecell}
\usepackage{array,multirow}
\usepackage{eurosym}

\expandafter\let\csname equation*\endcsname\relax
\expandafter\let\csname endequation*\endcsname\relax
\usepackage{amsmath}
\usepackage{amssymb}
\usepackage{url}

\usepackage[normalem]{ulem}
\usepackage{graphicx,epsfig}
\usepackage{dcolumn}
\usepackage{bm}
\usepackage{xcolor}

\def\cD{{\cal D}}
\def\cN{{\cal N}}

\def\cR{{\cal R}}

\newcommand{\req}[1]{Eq.~(\ref{#1})}
\newcommand{\avg}[1]{\langle #1\rangle}

\newcommand{\fig}[1]{Fig.~\ref{#1}}
\newcommand{\tab}[1]{Table \ref{#1}}

\DeclareMathOperator*{\argmin}{{\rm{argmin}}}

\makeatletter
\newcommand{\thickhline}{%
	\noalign {\ifnum 0=`}\fi \hrule height 1pt
	\futurelet \reserved@a \@xhline
}
\newcolumntype{"}{@{\hskip\tabcolsep\vrule width 2pt\hskip\tabcolsep}}
\makeatother

\newcommand{\cut}[1]{{}}
\usepackage{soul}
\newcommand{\spinspin}{\left<\Delta {\boldsymbol \sigma}^\mu \Delta {\boldsymbol \sigma}^\nu \right>}
\newcommand{\evt}{\mbox{$\tilde{E}^{V}$}}
\newcommand{\evtone}{\mbox{$\tilde{E}^{V}_{1}$}}
\newcommand{\evttwo}{\mbox{$\tilde{E}^{V}_{2}$}}
\newcommand{\evtj}{\mbox{$\tilde{E}^{V}_{j}$}}
\newcommand{\evtil}{\mbox{$\tilde{E}^{V}_{i\to l}$}}

\newcommand{\snij}{\sigma^\nu_{ij}}
\newcommand{\snji}{\sigma^\nu_{ji}}

\newcommand{\snijs}{\sigma^{\nu *}_{ij}}

\newcommand{\smijs}{\sigma^{\mu *}_{ij}}

\newcommand{\tsn}{\tilde{\sigma}^\nu}
\newcommand{\tsm}{\tilde{\sigma}^\mu}
\newcommand{\tsns}{\tilde{\sigma}^{\nu*}}
\newcommand{\tsms}{\tilde{\sigma}^{\mu*}}
\newcommand{\tsnij}{\tilde{\sigma}^\nu_{ij}}

\newcommand{\cH}{{\cal H}}
\newcommand{\cC}{{\cal C}}

\newcommand{\vs}{{\boldsymbol\sigma}}
\newcommand{\tvs}{{\tilde {\boldsymbol\sigma} }}
\newcommand{\vsss}{{\boldsymbol\sigma}^{*}}

\newcommand{\vsssgr}{{\boldsymbol\sigma}^{*}{(\gamma_r)}}
\newcommand{\tss}{{\tilde{\boldsymbol \sigma}}^*}
\newcommand{\tns}{{\tilde{\boldsymbol \sigma}}^{\nu *}}

\newcommand{\tn}{{\tilde{\boldsymbol \sigma}}^{\nu}}

\newcommand{\vns}{{\boldsymbol \sigma}^{\nu *}}
\newcommand{\vli}{{\bf \Lambda}_i}
\newcommand{\lione}{\Lambda_i^{\nu}}

\newcommand{\rione}{{R}_i^{\nu}}

\newcommand{\sjione}{\sigma_{ji}^{\nu}}

\newcommand{\Is}{I^*}

\newcommand{\D}{\Delta\cH}

\begin{document}
	\title{Re-routing game: The inadequacy of mean-field approach in modeling the herd behavior in path switching}
	\author{Ho Fai Po$^{1,2}$, Chi Ho Yeung$^1$}

	\affiliation{$^1$Department of Science and Environmental Studies, The Education University of Hong Kong, 10 Lo Ping Road, Taipo, Hong Kong. \\ $^2$Department of Mathematics, Aston University, B4 7ET, Birmingham, United Kingdom}
	\date{\today}


	\begin{abstract}
		Coordination of vehicle routes is a feasible way to ease traffic congestions amid a fixed road infrastructure. Nevertheless, even the optimal route configurations are provided to individual drivers, it is hard to achieve as greedy drivers may switch to other routes for a lower individual cost. Recent research uses mean-field cavity approach from the studies of spin glasses to analyze the impact of path switching in optimized transportation networks. However, this method only provides a mean-field approximation, which does not take into account the collective herd behavior in path switching due to un-coordinated individual decisions. In this study, we propose an exhaustive cavity approach to investigate the impact of un-coordinated path switching in a re-routing game and reveal that greedy drivers’ decision can be highly correlated which leads to the failure of mean-fielded approaches. Our theoretical results fits well with simulations, and our developed framework can be generalized to analyze other games with multiple players and rounds. Our results shed light on the impact of herd behavior of un-coordinated human drivers in suppressing congestions through path coordination.
	\end{abstract}
	
	\maketitle

	\section{Introduction}
	Traffic congestion is a severe problem in most metropolitan areas. It is costly in both monetary, societal and environmental terms. In Europe, traffic congestion costs nearly \euro 100 billion annually, which is about 1\% of the EU's GDP~\cite{urban-mobility}; while in US, congestions costed an average of 97 hours and \$1,348 per citizen in 2019~\cite{reed19}. Therefore, congestions are consuming money and time, let alone creating negative environmental impacts; easing traffic congestion is important and can potentially bring huge benefits to the society. However, relieving congestion through further infrastructure expansion is often unfeasible in established metropolitan areas, which leaves congestions recurrent and unresolvable. Coordinating traffic through globally optimized routes coupled with variable road-charges or financial inducements is one of the feasible and most promising approaches to alleviate congestion~\cite{Noh02,Dobrin01,Bayati08,yeung12,yeung2013physics,yeung2019coordinating}. 
	
	To achieve a global objective, some individuals often have to sacrifice and travel on a slightly longer path, despite the traveling time averaged over all vehicles will be reduced~\cite{yeung2013physics, yeung2019coordinating}. Therefore, even an optimally coordinated routes are provided, some vehicles may not follow which drive the system away from the optimum~\cite{shiftan11,prato09}. Different approaches are used to study individual greedy path switching~\cite{fischer04,anshelevich09, cole06, correa04}, while the impact of switching strategies on transportation networks is seldom discussed. Our recent work~\cite{po21} introduced a framework based on the cavity method~\cite{mezard87} to obtain a mean-field approximation of the impact incurred by path switching. This method successfully captures the trend and the features of transportation networks after path switching, but it can only provide a rough estimate of the numerical values of the quantities in the re-routed system. 
	
	 This inaccuracy is due to the fact that the mean-field approximation only captures the average behavior of road users in path switching, overlooking the potential correlation in their responses, which can lead to a large discrepancy from the real systems. As an example, the incomplete description by the mean-field approximation for the statistical inference in biological systems ~\cite{Terada_2019, Terada_2020}, where the assumption that coupling strengths are Gaussian distributed with zero mean and small variance does not hold, leading to a serious mismatch between the inferred and true structure. Instead, a more refined  analysis is needed to acquire reliable results\cite{po2024inferring}. Here, in modeling path switching, a more precise analysis is therefore necessary for understanding the impact of re-routing and herd behavior, which is crucial to the future development of intelligent transportation system via advanced information technology. 
	
	In this paper, we introduce a framework to exhaustively study the re-routing behavior of a small group of vehicles in the model of transportation network, i.e. a re-routing game introduced in~\cite{po21}. In this model, all vehicles are provided with routes which are globally and optimally coordinated, and some greedy vehicles do not follow the suggested routes and choose another route to minimize their own individual cost. Our method derived based on the cavity method in the studies of spin-glass~\cite{mezard87} exhaustively identifies the route choices of all vehicles, and reveals the impact of greedy re-routing on the whole system. The analytical results show that our exhaustive approach has an excellent agreement with simulations results. Remarkably, greedy vehicles can be highly correlated while taking un-coordinated switching decisions, leading to a herd effect and the failure of mean-field approach. Our work also demonstrates how the cavity method can be applied and generalized to study multi-agent game-theoretical problems with a small number of players exhaustively.
	
	The paper is organized as follows. We will introduce the model in Sec.~\ref{sec_model} and the theoretical derivation in Sec.~\ref{sec_ana_sol}. In Sec.~\ref{sec_results}, we compare the results obtained by the exhaustive and the mean-field cavity approach with simulations, and reveal the herd effect of path switching by studying the interaction between greedy vehicles. A summary is given in Sec.~\ref{sec_conclude}.
	
	\section{Model}
	\label{sec_model}
	
	We first describe the model, i.e. the re-routing game, to be studied; the same description was found in \cite{po21} since we will introduce our exhaustive cavity approach compared to the original mean-field approach introduced in \cite{po21}. Here, we consider $M$ vehicles denoted by $\nu=1,\dots,M$ traveling on a transportation network with $N$ nodes denoted as $ i = 1,2,3,\dots,N$, and denote ${\cal N}_i$ as the set of neighboring nodes of node $i$. The density of vehicles is denoted by $\alpha = M/N$. Each vehicle $\nu$ is traveling from a random origin node ${\cal O}_\nu$, and travels to a common destination node ${\cal D}$ that is randomly selected. We then define the movement of a vehicle $\nu$ on the link between node $i$ and $j$ as $\snij$, where
	\begin{align}
		\snij =
		\begin{cases}
			\pm1, & \mbox{if $\nu$ travels from $i\to j$ $(+)$ or $j \to i$ $(-)$,}
			\\
			0, &\mbox{if $\nu$ does not travel between $i$ and $j$,}
		\end{cases}
		\label{eq_sigma_ij}
	\end{align}
	and hence $\snij = \snji$. Therefore, the total volume of traffic flow from node $i$ to node $j$ is defined as $|I_{ij}|=\sum_\nu|\snij|$. To prevent traffic congestion due to the heavy load of roads usage, the \textit{social cost} per vehicle is introduced as 
	\begin{align}
		\cH(\vs|\gamma) = \frac{1}{M}\sum_{(ij)} \left( \sum_{\nu}|\snij| \right)^\gamma = \frac{1}{M}\sum_{(ij)}|I_{ij}|^\gamma,
		\label{eq_social_cost}
	\end{align}
	where $\gamma > 1$ and the cost increases faster than a linear function, penalizing the path overlap to ease traffic congestions; $(ij)$ denotes the un-ordered indices of the link connecting nodes $i$ and $j$; and the vector $\vs=\{\snij\}_{\nu}$ denotes the vector of path decisions for all vehicles $\nu$ and links $(ij)$. Noted that when $\gamma=1$, the two summations in \req{eq_social_cost} is interchangeable, and the social cost becomes
	\begin{align}
		\cH(\vs|\gamma=1) = \frac{1}{M}\sum_\nu\left(\sum_{ij}|\snij|\right),
		\label{eq_social_at_gamma_1}
	\end{align}
	meaning that all vehicles are taking the shortest path decisions, rather than considering the rest of the traffic condition. 
	
	It has been proven in~\cite{yeung12} that, when $\cH(\vs|\gamma)$ with $\gamma > 0$ is minimized, all vehicles traveling on the same link must go in the same direction, i.e. $\snij \geq 0$ or $\snij \leq 0$ for all $\nu$ on the link between node $i$ and $j$. Therefore, the directed volume of traffic flow between node $i$ and $j$ can be written as
	\begin{align}
		I_{ij} = \sum_\nu\snij, \mbox{and} \label{eq_sum_of_sig} \\
		\sum_\nu|\snij| = \left|\sum_\nu\snij\right| \label{eq_sum_of_sig}
	\end{align}
	where vehicles go from $i$ to $j$ if $I_{ij}>0$ and vice versa. Hence, the configuration of the recommended paths $\vs$ that minimizing $\cH(\vs|\gamma)$ can be found by employing the message-passing algorithm that will be discussed in Sec.~\ref{sec_ana_sol}.
	Now, we are going to establish the switching behaviors of greedy vehicles, over an optimized transportation network. We consider that the social cost $\cH(\vs|\gamma)$ is optimized, and the corresponding paths are recommended to all vehicles. The recommended path of vehicle $\nu$ is denoted as $\snijs=\pm1$ if the vehicle $\nu$ is suggested to pass from node $i$ to $j$ and from $j$ to $i$ respectively, and $\snijs=0$ otherwise. Hence, the vector of recommended paths is then
	\begin{align}
		\vsss = \argmin_\vs \cH(\vs|\gamma),
		\label{eq_recommend_paths}
	\end{align}
	where $\vsss = \{\snijs\}_{\nu,(ij)}$. Hence, the volume of the recommended traffic flow from node $i$ to $j$ is denoted as $\Is_{ij} = \sum_\nu\snijs$. Then, we consider that there exists a fraction $f_s$ of the $M$ vehicles are greedy, meaning that they do not follow the path recommendations if the path incurs a cost higher than their individual optimal cost. Instead, a greedy vehicle $\nu$ takes an alternative path, denoted as $\tn$, would minimize its own \textit{individual traveling cost} defined as
	\begin{align}
		\cC_\nu(\tn|\vsss,\gamma) = \sum_{(ij)}|\tsnij|\left(1+\left|\sum_{\mu\neq\nu}\smijs\right|\right)^{\gamma-1}.
		\label{eq_indi_cost}
	\end{align}
	This cost function of greedy vehicle $\nu$ implies that $\nu$ is taking the recommended volume of traffic induced by other vehicles, i.e.$\left|\sum_{\mu\neq\nu}\smijs\right|$, into consideration, and would like to minimize its overlap with the traffic. 
	The reason of the social cost and individual cost are defined as \req{eq_social_cost} and \req{eq_indi_cost} so that the sum of individual costs will becomes the social cost~\cite{po21}, i.e. $\sum_\nu \cC_\nu(\vs^{\nu *}|\vsss, \gamma) = \sum_\nu \sum_{i,j}|\snijs|\left|\sum_{\mu\neq\nu}\smijs\right|^{\gamma-1} = \sum_{i,j}\left|\sum_{\mu\neq\nu}\smijs\right|^\gamma = \cH(\vs|\gamma)$. This relation implies that \emph{the changes in social cost after path switching will only due to the changes of individual costs}, giving us an accurate comparison before and after path switching.
	
	To quantify the impact of selfish path switching behavior on the social cost, we measure the fractional change in social cost $\cH$ because of the path switching, defined as 
	\begin{align}
		\label{eq_delta}
		\D(\gamma_r, \gamma) = \frac{\cH(\tss(\gamma_r)|\gamma)-\cH(\vsssgr|\gamma)}{\cH(\vsssgr|\gamma)}
	\end{align}
	where $\vsssgr$ is the configuration of path recommended to the users which minimizes $\cH(\vs|\gamma_r)$; the social cost $\cH$ is characterized with parameter $\gamma$, not necessarily equal to $\gamma_r$; the vector $\tss(\gamma_r)$ is defined as the configuration of the updated path decisions, after the group of the greedy vehicles switch their paths, where,
	\begin{align}
		\tns(\gamma_r) 
		=
		\begin{cases}
			\vns(\gamma_r), &\mbox{for compliant users,} \nonumber
			\\
			\argmin_{\tn} \cC_\nu(\tn|\vsssgr, \gamma), &\mbox{for greedy users.} \nonumber
		\end{cases}
	\end{align}
	
	\subsection{Measurement of herd effect}
	
	To examine the existence of the herd effect of vehicles, we define a quantity to measure how likely the vehicles are traveling on the same link after switching. Consider two vehicles $\mu$ and $\nu$, we define the \textbf{\emph{conditional re-routing correlation}}, i.e. given that a flow of $I$ is initially found on the link, as 
 
	\begin{align}
		\label{eq_spinspin}
		\spinspin_{I^*}
		=
		\!\!\sum_{\{(ij)|I_{ij}=I^*\}}\sum_{\Delta { \sigma}^\mu_{ij}, \Delta { \sigma}^\nu_{ij}}
		\!\!\frac{ \Delta { \sigma_{ij}}^\mu \Delta { \sigma_{ij}}^\nu }{Z} 
		= 
		\!\!\sum_{\{ \tilde \sigma_{ij}^\mu, \tilde \sigma_{ij}^\nu, \sigma_{ij}^\mu, \sigma_{ij}^\nu \}}
		\!\!\frac{\left( |\tilde \sigma_{ij}^\mu| - |\sigma_{ij}^\mu| \right) \left( |\tilde \sigma_{ij}^\nu| - |\sigma_{ij}^\nu| \right)}{Z}
		, 
	\end{align}
	where $Z$ is the normalization constant given by 
	\begin{align}
		Z=\sum_{\{ \tilde \sigma_{ij}^\mu, \tilde \sigma_{ij}^\nu, \sigma_{ij}^\mu, \sigma_{ij}^\nu \}} (1-\delta_{|\tilde \sigma_{ij}^\mu|+|\sigma_{ij}^\mu| })(1-\delta_{|\tilde \sigma_{ij}^\nu|+|\sigma_{ij}^\nu| })\delta_{I_{ij},I^*}
	\end{align}
	This quantity is defined to include the interaction between two vehicles if they co-exist on the same link, such that empty links and links traveled by only one vehicle before and after re-routing are not included. The quantity $\Delta \sigma_{ij} = |\tilde \sigma_{ij}| - |\sigma_{ij}|$ can be understood as the change in route choices during re-routing , i.e. $\Delta \sigma_{ij} = 1$ means that a vehicle is initially not on the link $(ij)$ and switch to travel on this link after re-routing; $\Delta \sigma = -1$ means that a vehicle is initially on the link $(ij)$ and chooses not to travel on the link after re-routing, and; $\Delta \sigma = 0$ means that the vehicle does not change it decision (either traveling on the link or not) before and after re-routing. Therefore, $\spinspin_{I^*}$ measures the occurrence of vehicles $\mu$ and $\nu$ making the same switching choice on a link with initial flow$I=I^*$ before and after re-routing . Macroscopically, we can measure the herd effect by computing the \textbf{\emph{overall re-routing correlation}} as $\spinspin = \sum_I \spinspin_{I} P(I)$.
	
	\section{Analytical Solution}
	\label{sec_ana_sol}
	
	To study the interactions between vehicles, instead of measuring the macroscopic traffic condition using approaches in \cite{yeung12} and \cite{po21}, we develop a framework to exhaustively articulate the route choices of all individual vehicles. We adopt the two-stage optimization framework in~\cite{po21}, where the first stage identifies the path configurations that optimize the social cost $\cH(\vs|\gamma_r)$ and are recommended to all vehicles; while in the second stage, any greedy vehicle $\nu$ can then switch their path decisions to optimize their individual cost $\cC_\nu(\tn|\vsss,\gamma)$. To facilitate the analyses, we define the resource on node $i$ by $\vli = \{\Lambda_i^\nu\}_\nu$, where $\Lambda_i^\nu$ is the transportation load of the vehicle $\nu$ on node $i$, such that 
	\begin{align}
		\label{eq_lambda_i}
		\Lambda_i =
		\begin{cases}
			1, &\mbox{if $i = O_\nu$;}
			\\
			-\infty, &\mbox{if $i = \cD$;}
			\\
			0, &\mbox{otherwise.}
		\end{cases}
	\end{align}
	To ensure that every vehicle is assigned with a path traveling from their origin to the common destination, we define ${\bf R}_i = \{R_i^\nu\}_\nu$ as the net resource on each node $i \neq \cD$, satisfying
	\begin{align}
		\label{eq_Ri}
		\rione = \lione + \sum_{j \in {\cN}_i}\sjione = 0,
	\end{align}
	where $\cN_i$ is the set of neighboring nodes of $i$.
	
	We then employ the cavity approach~\cite{mezard87} at zero temperature and assume that only large loops exist; when node $i$ is removed, every neighbor of $i$ are approximately independent with each other. We define the optimized energy $E_{i\rightarrow l }(\vs_{il})$ of the tree network terminated at node $i$ as a function of the vector of routing decisions $\vs_{il}=\{\sigma_{il}^\nu\}_\nu$ for all vehicles $\nu$ from $i$ to $l$. Apart from measuring the total traffic flow~\cite{po21, yeung12}, this approach allows us to identify the contribution of energy by each vehicle on link $i$ to $l$. The recurrent relation relating $E_{i\rightarrow l }(\vs_{il})$ to the energies $E_{j \rightarrow i }(\vs_{ji})$ of its neighbors $j \in \cN_i$ excluding node $l$, is given by
	\begin{align}
		\label{eq_recur1}
		E_{i\rightarrow l }(\vs_{il}) = 
		\min_{\left\{ \{\vs_{ji}\}_{j\in\cN_i\setminus l} | {\bf R}_i={\bf0} \right\}}
		\!\left[ \left( \sum_\mu |\sigma_{il}^\mu| \right)^{\! \gamma_r} \!+\!\sum_{j\in \cN_i \setminus l}\! E_{j \rightarrow i }(\vs_{ji}) \right],
	\end{align}
	where $\gamma_r$ is the exponent responsible to optimize the social cost.
	
\begin{figure}
		\centerline{
			\includegraphics[width=0.6\linewidth]{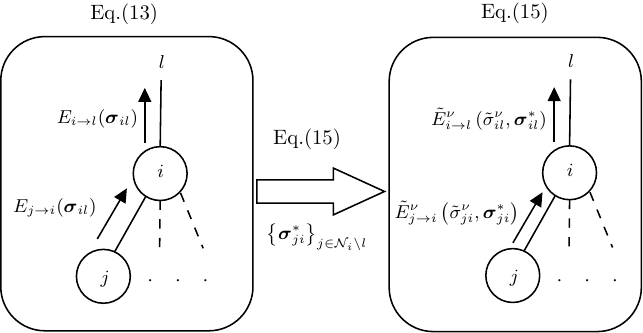} 
		}
		\caption{
			A diagram showing the two-stage optimization framework of the recursion relations in Eqs.~(\ref{eq_recur1}) - (\ref{eq_recur2}), and how (\ref{eq_recur2}) depends on (\ref{eq_recur1}).
		}
		\label{msgpassexplain}
	\end{figure}

	Next, to identify how vehicles switch, we first assume all users are greedy and introduce a set of energies $\{ {\tilde E}^\nu_{i\to l }(\tsn_{il}, \vsss_{il}) \}_\nu$ for all vehicles $\nu$, where each ${\tilde E}^\nu_{i\to l }(\tsn_{il}, \vsss_{il})$ represents the energy for $\nu$ to pick a new routing decision $\tsn_{il}$, after considering the flow in the recommended path configuration $\vsss_{il}$ of all vehicles that optimize the social cost $\cH(\vs|\gamma_r)$. Similar to \req{eq_recur1}, we can write down a recurrent relation for ${\tilde E}^\nu_{i\to l }(\tsn_{il}, \vsss_{il})$ at $i$ and the energies ${\tilde E}^\nu_{j\to i}(\tsn_{ji}, \vsss_{ji})$ of its neighbors $j \in \cN_i\setminus l$, given by
	{\small
		\begin{align}
			\label{eq_recur2}
			{\tilde E}^\nu_{i\to l }(\tsn_{il}, \vsss_{il}) = 
			\min_{ \left\{ \! \{\tsn_{ji}\}_{j\in\cN_i\setminus l} | {\bf R}_i={\bf0} \! \right\} } \! \left[ \left| \tsn_{il} \right| \! \left(\! 1 \! + \! \sum_{\mu \neq \nu} \left| \tsm_{il} \right| \right)^{\!\!\! \gamma - \!1} \!\!\!\!\! + \!\! \sum_{j\in \cN_i \setminus l}\!\!\! {\tilde E}^\nu_{j\to i }(\tsn_{ji}, \vsss_{ji}) \right] ,
		\end{align}
	}
	where $\gamma$ is the exponent responsible for optimizing the individual cost of each user $\nu$, and the quantity $\vsss_{ji}$ is the corresponding optimal decisions on edge $(ji)$, $\forall j \in \cN_i\setminus l$, provided that the optimal decision on the edge $(il)$ is $\vsss_{il}$. Using \req{eq_recur1} and given the vector $\vsss_{il}$, the set $\left\{ \vsss_{ji} \right\}_{j\in\cN_i\setminus l}$ can be computed by 
	\begin{align}
		\label{eq_argmin1}
		\left\{ \vsss_{ji} \right\}_{j\in\cN_i\setminus l} = 
		\argmin_{\left\{ \{\vs_{ji}\}_{j\in\cN_i\setminus l} | {\bf R}_i={\bf0} \right\}}
		\!\left[ \left( \sum_\mu |\sigma_{il}^\mu| \right)^{\! \gamma_r} \!+\!\sum_{j\in \cN_i \setminus l}\! E_{j \rightarrow i }(\vs_{ji}) \right].
	\end{align}
	To conclude, \fig{msgpassexplain} shows the two-stage message passing process: i) The social cost $\cH(\vs|\gamma_r)$ is minimized using \req{eq_recur1}. ii) The corresponding optimal individual decision $\vsss_{ji}, \forall j \in \cN_i\setminus l$, condition on the optimal social decision $\vsss_{il}$ is computed by \req{eq_argmin1}. iii) The set of individual costs of all vehicles $\{ {\tilde E}^\nu_{i\to l }(\tsn_{il}, \vsss_{il}) \}_\nu$ are computed using \req{eq_recur2} and the socially optimal path configuration in the first stage.
	We note that $E_{i\rightarrow l }(\vs_{il}) $ and $\{ {\tilde E}^\nu_{i\to l }(\tsn_{il}, \vsss_{il}) \}_\nu$ are extensive quantities dependent on the number of nodes in the network and are not suitable for testing the convergence of the analytical solutions. Therefore, we define the intensive quantities $E^V_{i\rightarrow l }(\vs_{il}) $ and $\{ \evtil^\nu(\tsn_{il}, \vsss_{il}) \}_\nu$ as 
	\begin{align}
		E^V_{i\rightarrow l }(\vs_{il}) & =   E_{i\rightarrow l }(\vs_{il}) - E_{i\rightarrow l }(\bf 0), \label{cR1} \\
		\evtil^\nu(\tsn_{il}, \vsss_{il}) & =  {\tilde E}^\nu_{i\to l }(\tsn_{il}, \vsss_{il}) - {\tilde E}^\nu_{i\to l }(0, \bf 0), \forall \nu, 
		\label{cR2}
	\end{align}
	where the computation of these quantities can be achievedby iterating Eqs.~(\ref{eq_recur1}) - (\ref{eq_recur2}).
	
	To obtain the analytical solution of the social cost and the detailed routing behavior of all vehicles before and after switching, we first need to find the joint functional probability distribution $P\left[ E^V(\vs), \left\{ \evt^\nu(\tsn, \vsss) \right\}_\nu \right]$, denoted as $P\left[ E^V , \left\{ \evt^\nu \right\}_\nu \right]$ for simplicity. Then, by using both \req{eq_recur1} and \req{eq_recur2}, we can write down the recurrent relation for $P\left[ E^V , \left\{ \evt^\nu \right\}_\nu \right]$ as 
	\begin{align}
		\label{eq_prob}
		P\left[ E^V , \left\{ \evt^\nu \right\}_\nu \right] = & \int dk \frac{P(k) k}{\avg{k}} \int d\Lambda P(\Lambda)
		\times \prod_{j=1}^{k-1} \int d E_j^V {\small \prod_{\mu=1}^M} d \evt^\mu P\left[ E^V , \left\{ \evt^\nu \right\}_\nu \right]
		\nonumber\\
		&\times \delta\!\left( E^V(\vs) -\cR_{16}\left[\{E_j^V\}, \Lambda, \vs\right]\right) \nonumber\\
		& \times\prod_{\mu=1}^M\delta\!\left(\evt^\mu(\tsm, \vsss)-\cR_{17}\left[\{\evtj^\mu\}, \Lambda, \tsm, \vsss\right]\right),
	\end{align}
	where $\cR_{16}$ and $\cR_{17}$ corresponds to the right hand side of the recurrent relations in \req{cR1} and \req{cR2} respectively. \req{eq_prob} can be then solved numerically for $P\left[ E^V , \left\{ \evt^\nu \right\}_\nu \right]$. 
	
	We remark that although the exhaustive cavity approach provides a detailed and accurate measurement for the social cost and configuration of paths as shown in Sec.~\ref{sec_results}, the computational complexity of this framework is much higher than the mean-field approach in~\cite{po21}. With $M$ as the number of vehicles, or the number of players if the current approach is applied to games, the complexity of the exhaustive approach is ${\cal O}\left( 3^{2M} \right)$, while the mean-field approach is ${\cal O}\left( \left(6M\right)^2\right)$, which is much less than the exhaustive approach. Thus, the mean-field approach is capable for anlyzing system with a large number of vehicles but only captures the general trend instead of precise re-routing behaviors such as the herd effect in switching, when players’ decisions are correlated. Therefore, the two approaches have different advantages and disadvantages and one can choose the appropriate one based on the need in analyses.
	
	\subsection{The probability of switching}
	
	With the joint probability distribution $P\!\!\left[\!E^V \!\!, \left\{ \!\! \evt^\nu \!\!\right\}_\nu \right]$, we can compute the probability $p(\tvs^*, \vsss)$ describing the joint probability distribution of the \emph{configurations of path switching} $\tvs^*$ and the \emph{optimal recommended traffic} $\vsss$ on a link, and this distribution allows us to measure the physical quantities desired, such as the conditional and the overall re-routing correlation in Eqs.~(\ref{eq_spinspin1}) and (\ref{eq_spinspin2}), respectively. The distribution $p(\tvs^*, \vsss)$ can be expressed as
	\begin{align}
		\label{eq_Preroute}
		p(\tvs^*, \vsss) = &\int dE_1^V \prod_\mu d\evtone^\mu P\left[ E_1^V , \left\{ \evtone^\nu \right\}_\nu \right] 
		\times \int dE_2^V \prod_\mu d\evttwo^\mu P\left[ E_2^V , \left\{ \evttwo^\nu \right\}_\nu \right] 
		\nonumber\\
		&\times\delta\left( \vsss -\argmin_{(\vs)}\left[E^V_1(\vs) + E^V_2(-\vs) - \left( \sum_\mu \left| \sigma^\mu\right| \right)^{\gamma_r}\right]\right)
		\nonumber\\
		&\times \prod_\mu \delta\!\!\left(\tilde{\sigma}^{\mu*} \!\!- \! \argmin_{\tilde{\sigma}^\mu}\!\!\left[ \! \evtone^\mu\!(\tsm, \vsss) \!+ \!\evttwo^\mu\!(\!-\tsm,\! -\vsss\!) \!- \!\left|\tilde{\sigma}^\mu\right|\!\!\left(\!1 \!\!+\!\! \sum_{\kappa\neq\mu} \!\left|\sigma^\kappa \right|\!\! \right)^{\!\!\!\!\gamma-1}\!\right]\!\!\right)\!,
	\end{align}
	which relies on the evaluation the energy shift in path switching.
	
	\subsection{The social cost after path switching}
	\cut{
		By marginalizing the joint distribution of $p(\tvs^*, \vsss)$, we can obtain the probabilities $p(\tvs^*)$ and $p(\vsss)$, which are the \emph{probability of path switching configurations for all vehicles} and the original recommended path configurations respectively, given by
		\begin{align}
			& p(\tvs^*) = \sum_{\vsss} p(\tvs^*, \vsss) , \label{eq_ptvs} \\
			& p(\vsss) = \sum_{\tvs^*} p(\tvs^*, \vsss) . \label{eq_pvsss} 
		\end{align}
	}
	By marginalizing the joint distribution of $p(\tvs^*, \vsss)$, we can obtain the probability of the recommended traffic on a link $p(I^*)$, given by
	\begin{align}
		\label{P_I_0}
		p(I^*) = \sum_{\tvs^*} \sum_{\vsss} p\left(\tvs^*, \vsss\right) \delta\left(\sum_\mu \left| \sigma^{\mu*} \right| - I^*\right).
	\end{align}
	Now, we consider that there are $Mf_s$ vehicles on the network that are greedy and define ${\tilde I}^*$ as the resulting traffic on a link after switching, then the probability $p({\tilde I}^*)$ can be evaluated by
	\begin{align}
		\label{P_I_new}
		p({\tilde I}^*) \! = \! \sum_{\tvs^*} \! \sum_{\vsss} p\!\left(\tvs^* \! , \!\vsss\! \right) \!\delta\!\!\left(\sum_{\mu=1}^{Mf_s} \left|\tsms\right| + \!\!\! \sum_{\mu=M\!f_s \! + \! 1}^{M} \!\! \! \left|\sigma^{\mu*}\right| \! - \! {\tilde I}^*\!\right)\!\!.
	\end{align}
	With \req{P_I_0} and \req{P_I_new}, we can compute the social cost $\cH(\vs^*|\gamma)$ and $\cH(\tvs^*|\gamma)$ before and after switching respectively, given by
	\begin{align}
		\cH(\vs^*(\gamma_r)|\gamma) = \sum_{I^*} p(I^*) I^*, \label{eq_ch}\\
		\cH(\tvs^*(\gamma_r)|\gamma) = \sum_{{\tilde I}^*} p({\tilde I}^*){\tilde I}^*. \label{eq_cht}
	\end{align}
	Then, the fractional change in social cost of the system, $\Delta\cH\left(\gamma_r, \gamma\right)$, can be evaluated by \req{eq_delta}.
	
	\subsection{Re-routing correlation}
	
	To investigate the herd effect, we consider the changes between routing configurations of two vehicles $\mu$ and $\nu$ before and after re-routing. We first define the joint distribution $p(\tsn, \tsn, \sigma^\mu, \sigma^\nu, I^*)$ as the probability of the path choices of $\mu$ and $\nu$ before and after re-routing on a link with original optimized flow $I^*$, which can be computed by
	\begin{align}
		\label{eq_tilsig12}
		p(\tsm, \tsn, \sigma^\mu, \sigma^\nu, I^*) = & \sum_{\tvs^*} \sum_{\vsss} p\left(\tvs^*, \vsss\right) \delta\!\!\left(\!\sum_\mu \left| \sigma^\mu \right| \! - \! I^*\!\!\right) \nonumber \\
		& \times \delta(\tsm - \tsms) \delta(\tsn - \tsns) 
		\times \delta(\sigma^\mu - \sigma^{\mu*} ) \delta(\sigma^\nu - \sigma^{\nu*} ) .
	\end{align}
	Thus, with $p(\tsm, \tsn, \sigma^\mu, \sigma^\nu | I^*) = p(\tsm, \tsn, \sigma^\mu, \sigma^\nu, I^*)/p(I^*)$, we can compute the \emph{conditional re-routing correlation}, given by
	\begin{align}
		\label{eq_spinspin1}
		\spinspin_{I^*} = 
		\sum_{ \{ \tsm , \tsn, \sigma^\mu, \sigma^\nu \} } \!\!\!\!\!\!\!\!\! \left(\left| \tsm \right| - \left| \sigma^\mu \right|\right) \left(\left| \tsn \right| - \left| \sigma^\nu \right|\right) p\left(\tsm, \tsn, \sigma^\mu, \sigma^\nu \left| I^* \right.\right),
	\end{align}
	and the \emph{overall re-routing correlation} is given by
	\begin{align}
		\label{eq_spinspin2}
		\spinspin = \sum_{I^*} \spinspin_{I^*} p(I^*).
	\end{align}
	We note that the possible values of $\Delta{\bf \sigma}$ are 1,0 and -1, hence, the correlations $\spinspin_{I^*}$ and $\spinspin$ are bounded between -1 and 1. These quantities characterize the correlations of vehicles in making the same re-routing decision on a link. Negative $\spinspin$ suggests that the two vehicles make an opposite decision, for instance, $\mu$ initially on the link left the link after re-routing, while $\nu$ initially not on the link switches to travel via the link after re-routing. On the other hand, positive $\spinspin$ suggests that the two vehicles make the same re-routing decision, for instance, they switch to travel on or out of the link together.
	
	\section{Results}
	\label{sec_results}
	
	To illustrate that our exhaustive cavity approach provides an accurate analytic solution to the path switching behavior, we focus on two scenarios where the initial path configurations are computed based on $\gamma_r = 1 \mbox{ and } 2$, and individual vehicles then re-route accouding to $\gamma = 2$. Due to the high computational complexity of this exhaustive cavity approach, we examine transportation networks with a different density $\alpha=M/N$ by fixing the number of vehicles to be $M=8$ and vary the number of nodes $N$ . We then compare the analytic solutions obtained by our exhaustive cavity approach with those obtained by the mean-field approach developed in \cite{po21} as well as simulation results. 
	
	As shown in \fig{fig_frac_deltaE}(a) and (b), for both scenarios of $(\gamma_r, \gamma)=(1,2)$ and $(2,2)$, our exhaustive cavity approach agrees much better with simulation results than the mean-field approach in terms of the fractional change in travel cost $\D$. The mean-field approach yields a smaller value of $\D$ compared to simulations but still capture the trend. This suggests that if one prefers accurate analytical results describing the re-routing behavior, the exhaustive cavity approach which articulates all route choices in \req{eq_recur1} and \req{eq_recur2} for a small group of vehicles is a better option, compared to the mean-field approach which is capable to analyze a system with a larger number of vehicles.
	
	In particular, our exhaustive cavity approach well predict the critical density of re-routing users, denoted as $f_s^*$, beyond which the gain in the social cost of the system ceases in the case of $\gamma=1$. In \tab{tab_critical_fs}, we show the critical density $f_s^*$ obtained from simulations, our exhaustive vacity approach, as well as the mean-field approach, over different network density $\alpha$. As we can see from the table, $f_s^*$ obtained using our exhaustive cavity approach is closer than that obtained by the mean-field approach, in comparison to  the simulation results. In addition, since mean-field approach overestimates the gain in social cost because the correlated path-switching decisions are not considered, its computed $\Delta \cH$ tends to be more negative; when $\alpha$ is large, i.e. there are more vehicles, $\Delta \cH<0$ for all $f_s$, and $f_s^*$ only exist when $\alpha$ is small, and is far away from the simulation results. The results suggest that compare to the mean-field approach, our exhaustive cavity approach performs better in identifying critical density of re-routing users when gain ceases to exist.
	
	\begin{table}
		\begin{center}
			\begin{tabular}{|c||c|c|c|}
				\hline 
				$\alpha$ & $f_s^*$(Simulation) & $f_s^*$(Exhaustive) & $f_s^*$(Mean-Field) \tabularnewline
				\hline 
				\hline 
				0.1 & 0.76 & 0.79 & 0.97\tabularnewline
				\hline 
				0.2 & 0.75 & 0.76 & 0.94\tabularnewline
				\hline 
				0.3 & 0.72 & 0.76 & NA\tabularnewline
				\hline 
				0.4 & 0.72 & 0.77 & NA\tabularnewline
				\hline 
				0.5 & 0.71 & 0.78 & NA\tabularnewline
				\hline 
				0.8 & 0.71 & 0.81 & NA\tabularnewline
				\hline 
			\end{tabular}å
		\end{center}
		\caption{The critical density of selfish users $f_s^*$ beyond which the gain in the social cost of the system ceases in the case of $\gamma=1$, over different values of the network density $\alpha$.}
		\label{tab_critical_fs}
	\end{table}

	\subsection{Herd effect in re-routing}
	
	The discrepancies between the exhaustive and the mean-field approaches imply that the latter cannot capture some fundamental mechanism in route switching. We first note that in the mean-field approach, the traffic condition after re-routing is estimated by computing the re-routing choice of a single specific vehicle, which is used for estimating and representing the mean-field switching probability for all vehicles, assuming all of them make an independent re-routing decision. The discrepancies thus come from the invalidity of this mean-field assumption, suggesting a correlation underlying re-routing decisions of individual vehicles, for instance, a route favorable for most vehicles, which can be considered as a herd effect in path switching.
	
	\begin{figure}
		\centerline{
			\epsfig{figure=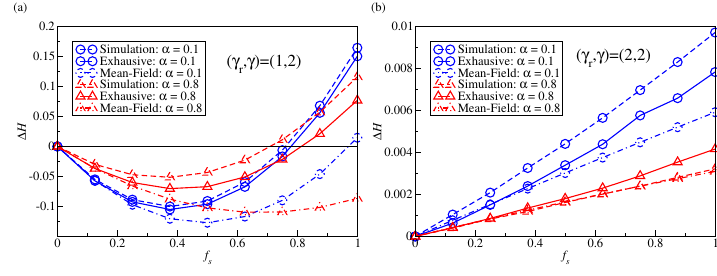, width=0.9\linewidth}
		}
		\caption{
			The fractional change in social costs $\Delta\cH$ as a function of $f_s$, averaged over all vehicles for: $(\gamma_r, \gamma)=(1,2)$ in (a) and; $(\gamma_r, \gamma)=(2,2)$ in (b). The simulation results are obtained by averaging 1000 realizations, on random regular graphs with a fixed number of vehicles $M=8$ and $k=3$ for density $\alpha=M/N=0.1 \mbox{ and } 0.8$. Analytical results using the mean-field cavity approach\cite{po21} and the exhaustive cavity approach are shown for comparison. 
		}
		\label{fig_frac_deltaE}
	\end{figure}
	
\begin{figure}
		\centerline{
			\epsfig{figure=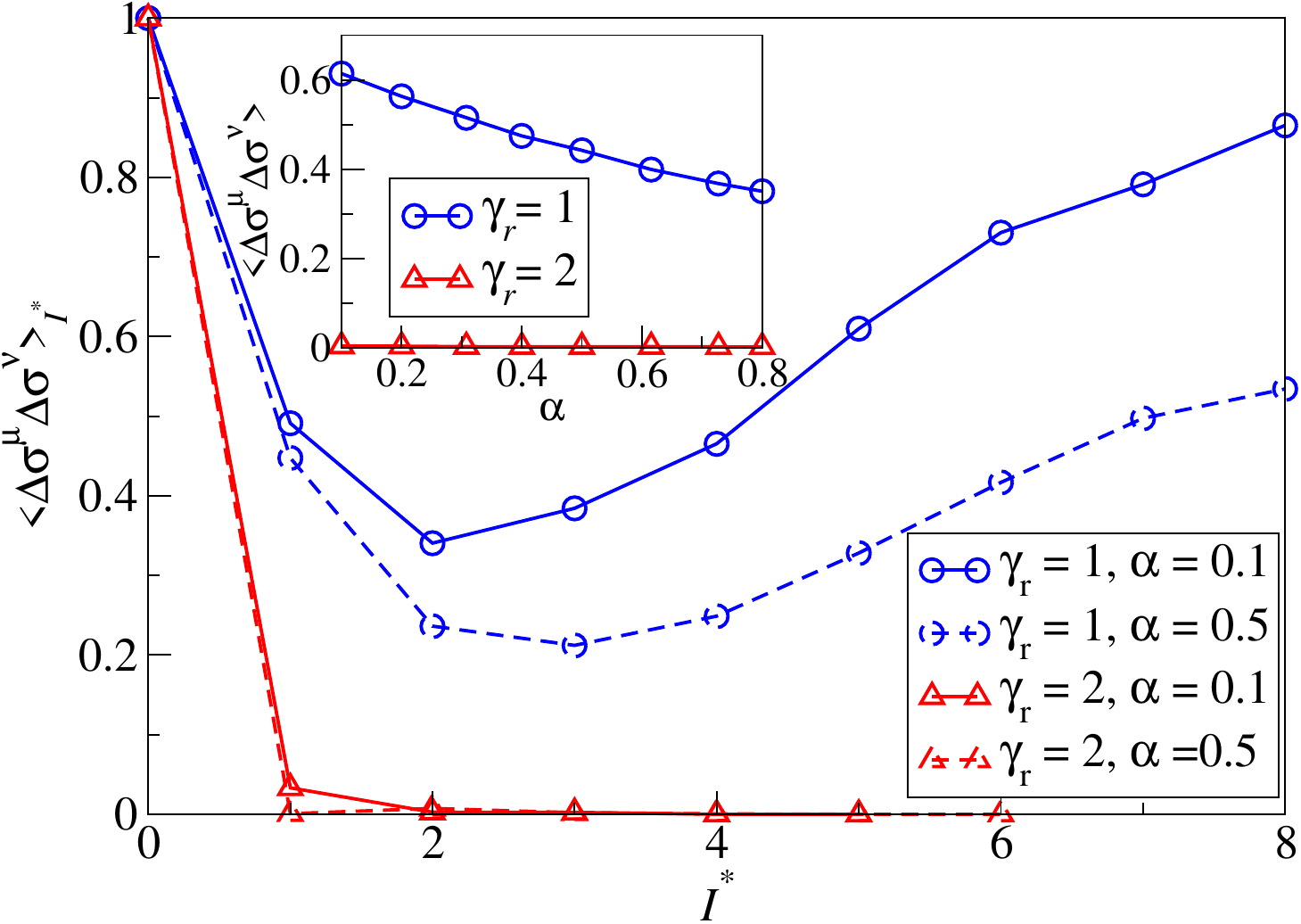, width=0.5\linewidth}
		}
		\caption{
			The conditional spin-spin correlation of routing, $\spinspin_{I_0}$ as a function of $I_0$, for $\gamma_r=1\mbox{ and }2$, $\alpha=0.1\mbox{ and }0.5$. insets: The spin-spin correlation of routing $\spinspin$ as a function of $\alpha$, for $\gamma_r=1\mbox{ and }2$. The results are the analytical solutions obtained by the exhaustive cavity approach.
		}
		\label{fig_spinspin}
	\end{figure}

	To examine the herd behavior in re-routing, we compute the conditional and the overall re-routing correlation, i.e. $\spinspin_{I^*}$ and $\spinspin$ in \req{eq_spinspin1} and \req{eq_spinspin2}, as shown in \fig{fig_spinspin} and its inset respectively. As we can see, when $\gamma_r=2$, where path configurations optimizing the social cost are initially recommended to all vehicles, the values of $\spinspin_{I^*}$ and $\spinspin$ are very small for all $\alpha$ and $I^*\neq1$. We also note that when $I^* = 0$, $\spinspin=1$ since $\left| \sigma^* \right|=0$ and hence $|\tilde \sigma| = 1$ if $|\tilde \sigma|+|\sigma^*| \neq 0$. This suggests that over a globally optimized transportation network, when traffic flows tend to be evenly distributed on the whole network, roads that are favorable to most vehicles do not exist and hence, greedy vehicles tend to switch in an un-correlated manner. This also explains that when $\gamma_r=2$, $\D$ obtained by the mean-field approach gives a better estimation than that obtained when $\gamma_r=1$, as shown in \fig{fig_frac_deltaE}, since re-routing is less correlated in the former.
	
	On the other hand, in the case of $\gamma_r=1$ with $\alpha$ = 0.1 or 0.5, $\spinspin_{I^*} > 0$ for all $I^*$ as shown in \fig{fig_spinspin}. In particular, $\spinspin_{I^*}$ first decreases then increases as $I^*$ increases. When the initially recommended traffic flow $I^*$ on a road is small, $\left| \sigma^* \right|\approx 0$ and $\Delta\sigma = \left| \tilde \sigma \right| - \left| \sigma^* \right| > 0$ implies $\left| \tilde \sigma \right| >0$ in most cases, suggesting the two vehicles are switching into a road that they initially do not travel on. Therefore, $\spinspin_{I^*} > 0$ implies that greedy vehicles consider the road to be relatively empty and favorable to ride on, such that they switch to the road together. Then when $I^*$ increases, $\spinspin_{I^*}$ decreases as the switching options become more even, and greedy vehicles can either stay or leave the recommended road to minimize their individual costs. Finally, when $I^*$ further increases, since the shortest paths are recommended to vehicles in the case of $\gamma_r=1$, almost all vehicles initially travel on a relatively small set of roads, the greedy vehicles can choose to stay ($\Delta\sigma = 0$) or leave ($\Delta\sigma = 1$) the road. An increasing $\spinspin_{I^*}$ therefore implies that more greedy vehicles choose to leave the road together, leading to a herd effect in switching. 
	
	Next, we examine the dependence of herd behavior on vehicle density. As shown in the inset of \fig{fig_spinspin}, the overall re-routing correlation $\spinspin$ decreases as the density of vehicles $\alpha$ increases. This can be understood as, when the vehicle density is small, most vehicles are recommended to travel on a limited number of paths, leading to under-loaded roads and multiple greedy vehicles tend to switch to these roads which lead to a relatively higher $\spinspin$ than the case with high vehicle density $\alpha$.
	The above results show that there is a high correlation when vehicles re-route, lead to a herb behavior and thus sub-optimal route configurations.
	
	\section{Conclusion}
	\label{sec_conclude}
	In this study, we studied a two-step transportation re-routing game based on the mathematical model established in \cite{po21}. Initially, every vehicle is provided with a recommended route from their starting point to a universal destination, which optimizes the social cost if all of them follow the recommendation. However, greedy vehicles might switch their paths to minimize their individual traffic cost based on the recommendation, causing the cost to deviate from the original globally optimized social cost. An exhaustive cavity approach is developed to analyze the path switching properties of greedy vehicles, especially the correlation in their re-routing decisions. We found that the exhaustive cavity approach leads to analytical results agree better with simulation compared to a mean-field version of the cavity approach.
	
	 Nevertheless, although the introduced exhaustive cavity approach is more accurate, its computational complexity is higher compared to the mean-field approach. Therefore, it demonstrate a trade-off between complexity and accuracy, as in many other analytical examples.. On top of that, we revealed the reason behind the mismatch between the mean-field  approach and the simulation results, which is the herd effect in path switching, when vehicles’ re-routing decisions are correlated.. In our exhaustive cavity approach, we are able to compute the conditional re-routing correlation, which is the essence of the approach in outperforming the mean-field estimation.

In terms of the re-routing game, by using the exhaustive cavity approach, we showed that greedy vehicles’ re-routing decisions are highly correlated, i.e. likely to re-route to the same path, in the case of an uncoordinated transportation network; while vehicles are almost uncorrelated under on the other hand, in the case of coordinated and optimized networks, vehicles’ re-routing decisions tend to be uncorrelated. In a broader context, our study highlights that  the mean-field approximation may fail when correlation emerges, which is consistent with observations in many other applications of mean-field approach.
	
	Our new theoretical framework provide insights for us to improve mean field cavity approach. For instead, in the mean-field cavity approach, instead of single out one vehicle or exhaustively account for all vehicles, one can extract two vehicles and compute the re-routing correlation for these two vehicles. Our introduced approach can also be used to improve the existing navigation systems for a more precise recommendation of routes and the prediction of traffic condition.. Our work also shows how cavity method can be used to analyze a multi-step optimization problem exhaustively. The generalized theoretical framework can be applied to study other problems that based on games with multiple steps and involving multi-players that are responding to state of the system.

	\section*{Acknowledgments}
	\bibliographystyle{prsty}
	\bibliography{exhaustivecite}

\begin{thebibliography}{10}

\bibitem{urban-mobility}
{Directorate-General for Mobility and Transport}, Clean transport, Urban
  transport, Urban mobility,
  \url{https://ec.europa.eu/transport/themes/urban/urban_mobility_en}, 2021.

\bibitem{reed19}
T. Reed,   (2019).

\bibitem{Noh02}
J.~D. Noh and H. Rieger, Phys. Rev. E {\bf 66},  066127  (2002).

\bibitem{Dobrin01}
R. Dobrin and P.~M. Duxbury, Phys. Rev. Lett. {\bf 86},  5076  (2001).

\bibitem{Bayati08}
M. Bayati {\it et~al.}, Phys. Rev. Lett. {\bf 101},  037208  (2008).

\bibitem{yeung12}
C.~H. Yeung and D. Saad, Physical review letters {\bf 108},  208701  (2012).

\bibitem{yeung2013physics}
C.~H. Yeung, D. Saad, and K.~M. Wong, Proceedings of the National Academy of
  Sciences {\bf 110},  13717  (2013).

\bibitem{yeung2019coordinating}
C.~H. Yeung, Physical Review E {\bf 99},  042123  (2019).

\bibitem{shiftan11}
Y. Shiftan, S. Bekhor, and G. Albert, IET intelligent transport systems {\bf
  5},  183  (2011).

\bibitem{prato09}
C.~G. Prato, Journal of choice modelling {\bf 2},  65  (2009).

\bibitem{fischer04}
S. Fischer and B. V{\"o}cking,  in {\em European Symposium on Algorithms},
  Springer (PUBLISHER, ADDRESS, 2004), pp.\ 323--334.

\bibitem{anshelevich09}
E. Anshelevich and S. Ukkusuri,  in {\em International Symposium on Algorithmic
  Game Theory}, Springer (PUBLISHER, ADDRESS, 2009), pp.\ 171--182.

\bibitem{cole06}
R. Cole, Y. Dodis, and T. Roughgarden, Journal of Computer and System Sciences
  {\bf 72},  444  (2006).

\bibitem{correa04}
J.~R. Correa, A.~S. Schulz, and N.~E. Stier-Moses, Mathematics of Operations
  Research {\bf 29},  961  (2004).

\bibitem{po21}
H.~F. Po, C.~H. Yeung, and D. Saad, Physical Review E {\bf 103},  022306
  (2021).

\bibitem{mezard87}
M. M{\'e}zard, G. Parisi, and M. Virasoro, {\em Spin glass theory and beyond:
  An Introduction to the Replica Method and Its Applications} (World Scientific
  Publishing Company, ADDRESS, 1987), Vol.~9.

\bibitem{Terada_2019}
Y. Terada, T. Obuchi, T. Isomura, and Y. Kabashima, Journal of Statistical
  Mechanics: Theory and Experiment {\bf 2019},  124010  (2019).

\bibitem{Terada_2020}
Y. Terada, T. Obuchi, T. Isomura, and Y. Kabashima, Neural Computation {\bf
  32},  2187  (2020).

\bibitem{po2024inferring}
H.~F. Po {\it et~al.}, Inferring Structure of Cortical Neuronal Networks from
  Firing Data: A Statistical Physics Approach, 2024.

\end{thebibliography}
\end{document}